\documentclass[sn-mathphys-num]{sn-jnl}

\usepackage{graphicx}%
\usepackage{multirow}%
\usepackage{amsmath,amssymb,amsfonts}%
\usepackage{amsthm}%
\usepackage{mathrsfs}%
\usepackage[title]{appendix}%
\usepackage{xcolor}%
\usepackage{textcomp}%
\usepackage{manyfoot}%
\usepackage{booktabs}%
\usepackage{upgreek}
\usepackage{algorithm}%
\usepackage{algorithmicx}%
\usepackage{algpseudocode}%
\usepackage{listings}%
\usepackage[latin2]{inputenc}
\usepackage{xspace}
\usepackage{lineno}

\newcommand*{\sqs}{\ensuremath{\sqrt{s}}\xspace}
\newcommand*{\sqsn}{\ensuremath{\sqrt{s_{\rm NN}}}\xspace}

\newcommand*{\pT}{\ensuremath{p_\mathrm{T}}\xspace}

\newcommand*{\MeV}{\ensuremath{\mathrm{MeV}}\xspace}

\newcommand*{\Lcp}{\ensuremath{\Lambda_\mathrm{c}^{+}}\xspace}
\newcommand*{\Dz}{\ensuremath{{\rm D}^{0}}\xspace}
\newcommand*{\Dsp}{\ensuremath{{\rm D}^{\ast+}}\xspace}

\begin{document}

\title[The ALICE 3 detector concept]{The ALICE 3 detector concept for LHC Runs 5 and 6 and its physics performance}

\author*{\fnm{Róbert} \sur{Vértesi}}\email{vertesi.robert@wigner.hu}
\equalcont{for the ALICE Collaboration}

\affil{\orgdiv{Department of High Energy Physics}, \orgname{HUN-REN Wigner Research Centre for Physics}, \orgaddress{\street{29-33 Konkoly-Thege Miklós út}, \city{Budapest}, \postcode{1121}, \country{Hungary}}}

\abstract{The LHC Run 5 and 6 data taking phases will bring unprecedented luminosity in high-energy proton-proton and in heavy-ion collisions. The ALICE Collaboration proposes a next-generation experiment, ALICE 3, specifically designed to operate with the future LHC. ALICE 3 will feature a large pixel-based tracking system covering eight units of pseudorapidity, complemented by advanced particle identification systems. These include silicon time-of-flight layers, a ring-imaging Cherenkov detector, a muon identification system, and an electromagnetic calorimeter. By placing the vertex detector on a retractable plate inside the beam pipe, a track pointing resolution better than 10 microns can be achieved for the transverse momentum range $\pT>200\ \MeV /c$.
ALICE 3 will be capable of innovative measurements of the quark-gluon plasma (QGP) and explore new frontiers in quantum chromodynamics (QCD). The detailed study of thermal and dynamical properties of QGP  will be made possible by measuring low-\pT heavy-flavour production, including beauty hadrons, multi-charm baryons, and charm-charm correlations. Precise multi-differential measurements of dielectron emission will allow for the exploration of chiral-symmetry restoration and the time-evolution of QGP temperature. In addition to QGP studies, ALICE 3 will make unique contributions to the physics of the hadronic phase, through femtoscopic studies of charm meson interaction potentials and searches for nuclei containing charm. This contribution covers the detector design, expected physics performance, and the current status of detector research and development.}


\maketitle

\section{Introduction}\label{intro}

In ultra-relativistic heavy-ion collisions, hadronic matter transforms into a deconfined state of quarks and gluons, called the quark-gluon plasma (QGP). The primary objective of ALICE (A Large Ion Collider Experiment) is to explore the dynamics of this extreme state of the strongly interacting matter. Results from the LHC Runs 1 and 2 have significantly expanded our comprehension of the QGP and the QCD phase diagram~\cite{ALICE:2022wpn}. With the upgrades brought by ALICE 2, and those of the forthcoming ALICE 2.1, systematic studies during LHC Runs 3 and 4 will further enhance our understanding of the fundamental properties of the QGP. However, despite these advancements, several critical questions will remain open beyond Run 4. These include understanding the equilibration process of the QGP, the partonic equation of state and its dependence on the temperature, the dynamics of chiral symmetry restoration, as well as the hadronization mechanisms of the deconfined matter. To address these questions, precision differential measurements of dileptons, systematic studies of (multi-)heavy-flavored hadrons at low transverse momentum, and measuring hadronic interactions are essential. These investigations require significant improvements in detector performance and data collection capabilities. ALICE 3, a new detector concept planned to replace ALICE for Runs 5 and 6, will provide the required capabilities and make significant contributions to answering these questions~\cite{ALICE:2022wwr}.

\section{Design and performance}

\begin{figure}[h]
	\centering
	\begin{minipage}{.49\textwidth}
		\centering
		\includegraphics[width=\linewidth]{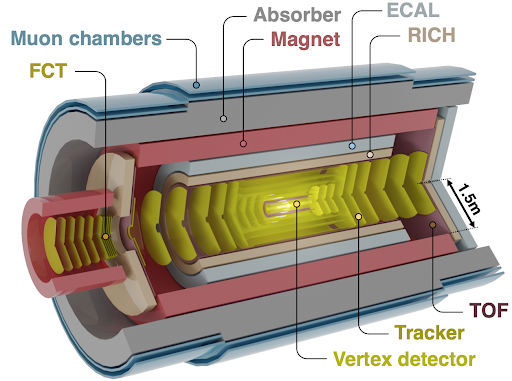}
		\caption{A schematic drawing of the ALICE 3 detector system.}
		\label{fig:alice3}
	\end{minipage}
	\hfill
	\begin{minipage}{.49\textwidth}
		\centering
		\includegraphics[width=\linewidth]{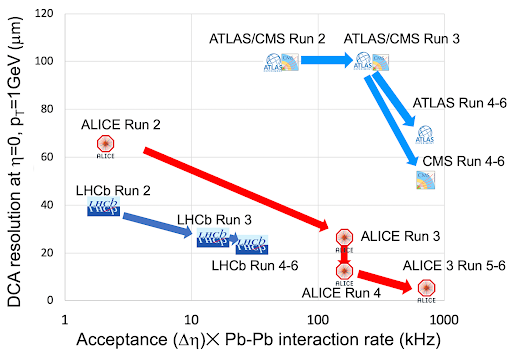}
		\caption{Effect of planned upgrades on the performance of the LHC experiments.}
		\label{fig:dca-accXrate}
	\end{minipage}
\end{figure}
The ALICE 3 concept incorporates cutting-edge technologies to meet its ambitious physics goals, featuring a broad pseudorapidity coverage of $|\eta|<4$. Its schematics and basic performance requirements are shown in Figs.~\ref{fig:alice3} and \ref{fig:dca-accXrate}, respectively. The design of a compact silicon tracker composed of vertex detector placed on a retractable plate in vacuum inside the beam pipe, and an outer tracker, are both based on Monolithic Active Pixel Sensor technology. 
\begin{figure}[h]
	\centering
	\begin{minipage}{.42\textwidth}
		\centering
		\includegraphics[width=\linewidth]{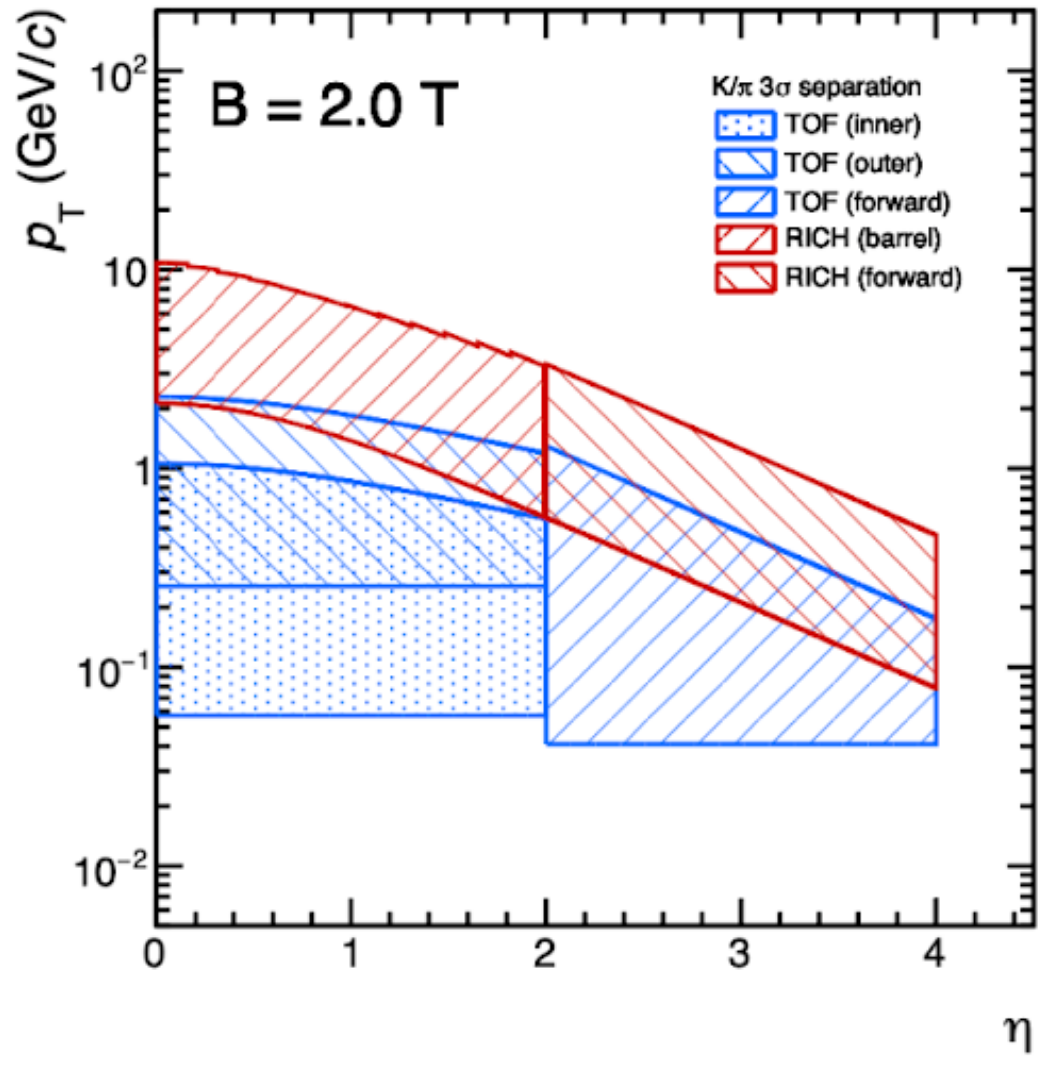}
		\caption{Combined charged-kaon--pion separation capabilities of TOF and RICH.}
		\label{fig:tof_rich_Kpisep}
	\end{minipage}
	\hfill
	\begin{minipage}{.56\textwidth}
		\centering
		\includegraphics[width=\linewidth]{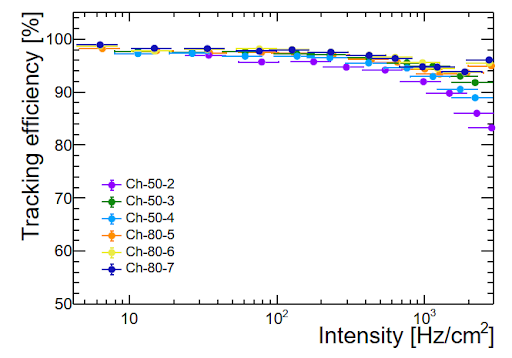}
		\caption{Detection efficiency of the MWPC for the muon identification system, as a function of particle flux density, from test beam measurements with a beam momentum of 5 GeV/$c$.~\cite{Alfaro:2024sxc}}
		\label{fig:mwpc_eff}
	\end{minipage}
\end{figure}
The vertex detector consists of 3 layers and 3 disks on each side, having an excellent pointing resolution of of $3$--$4$ $\upmu$m and a very low material budget of $\sim$0.1\% $X_0$ per layer. The outer tracking system of 8 layers and 9 disks on each side, corresponds to $\sim$1\% $X_0$ per layer. Tracking is facilitated by a superconducting magnet capable of generating a magnetic field of 2 T. Particle identification is carried out by the silicon-based time-of-flight detector of 2 layers and 2 forward disks, complemented by an aerogel ring-imaging Cherenkov detector. Their combined charged-particle identification capabilities are demonstrated in Fig.~\ref{fig:tof_rich_Kpisep}. The identification of electrons and photons are facilitated by an electromagnetic calorimeter, comprising a sampling lead-scintillator detector of a barrel layer and a disk layer, and a high-resolution PbWO$_4$ crystal segment at $|\eta|<0.22$ for precise single-electron measurements. A muon identification detector will be located outside a hadron absorber, approximately $\sim$70 cm thick, made of steel. Several technology options are being considered~\cite{Alfaro:2024sxc}, with a muon detection efficiency design goal of at least 98\% up to a particle flux density of 10 Hz/cm$^2$ (Fig.~\ref{fig:mwpc_eff}).

\section{Physics capabilities}

The ALICE 3 detector is planned to complete a rich physics program to explore the quark-gluon plasma properties, hadron-hadron interactions, and search for physics beyond the Standard Model~\cite{ALICE:2022wwr}. Some examples are described in the following.

\begin{figure}[h]
	\centering
	\begin{minipage}{.48\textwidth}
		\centering
		\includegraphics[width=\linewidth]{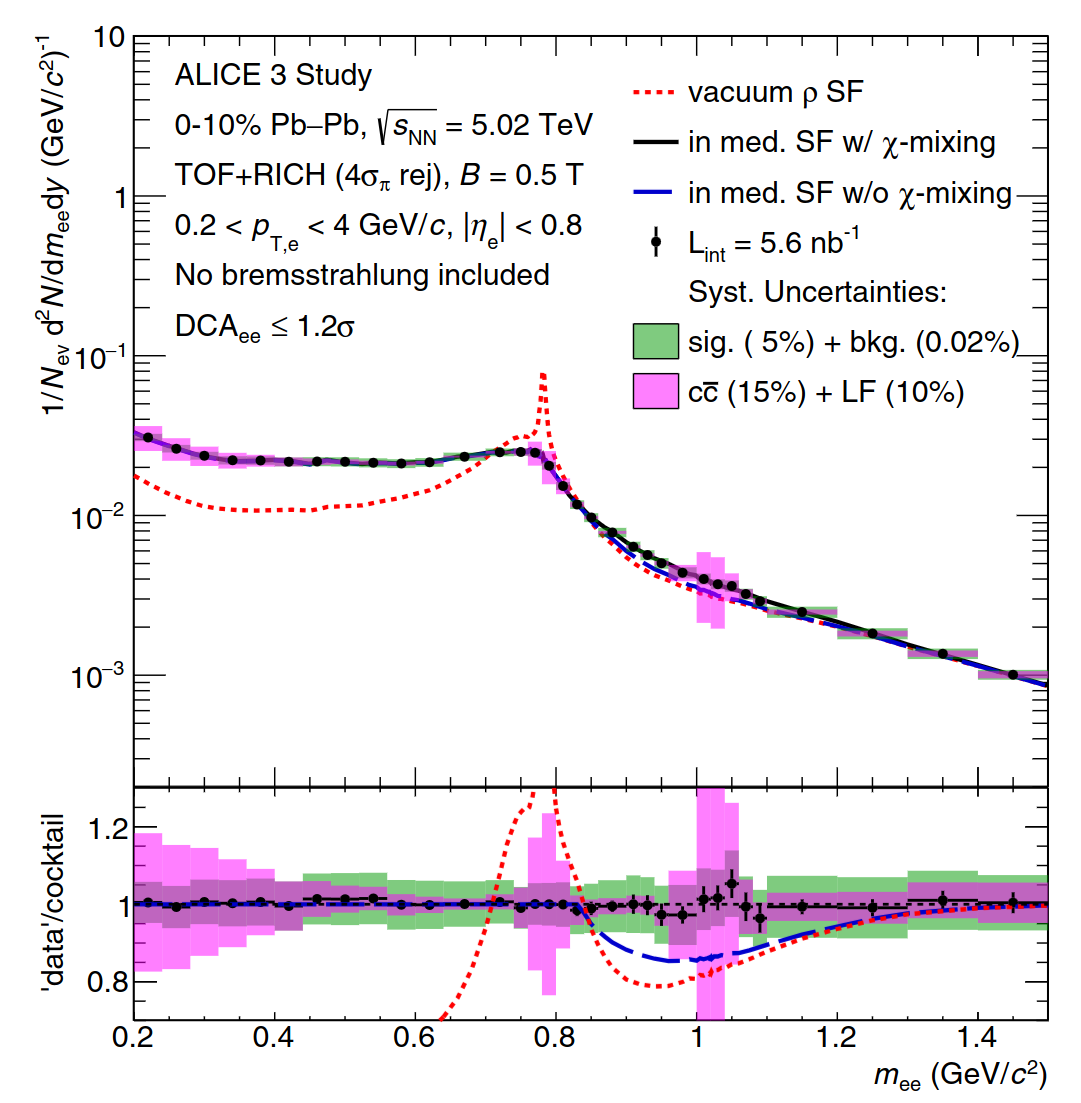}
		\caption{Simulated excess ${\rm e}^{+}{\rm e}^{-}$ pairs after subtraction of correlated light-hadron and heavy-flavour hadron decays at midrapidity in central Pb--Pb collisions at \sqsn= 5.02 TeV, compared to predictions with modified and vacuum $\rho$ spectral functions.}
		\label{fig:ee_lowmass}
	\end{minipage}
	\hfill
	\begin{minipage}{.48\textwidth}
		\centering
		\includegraphics[width=\linewidth]{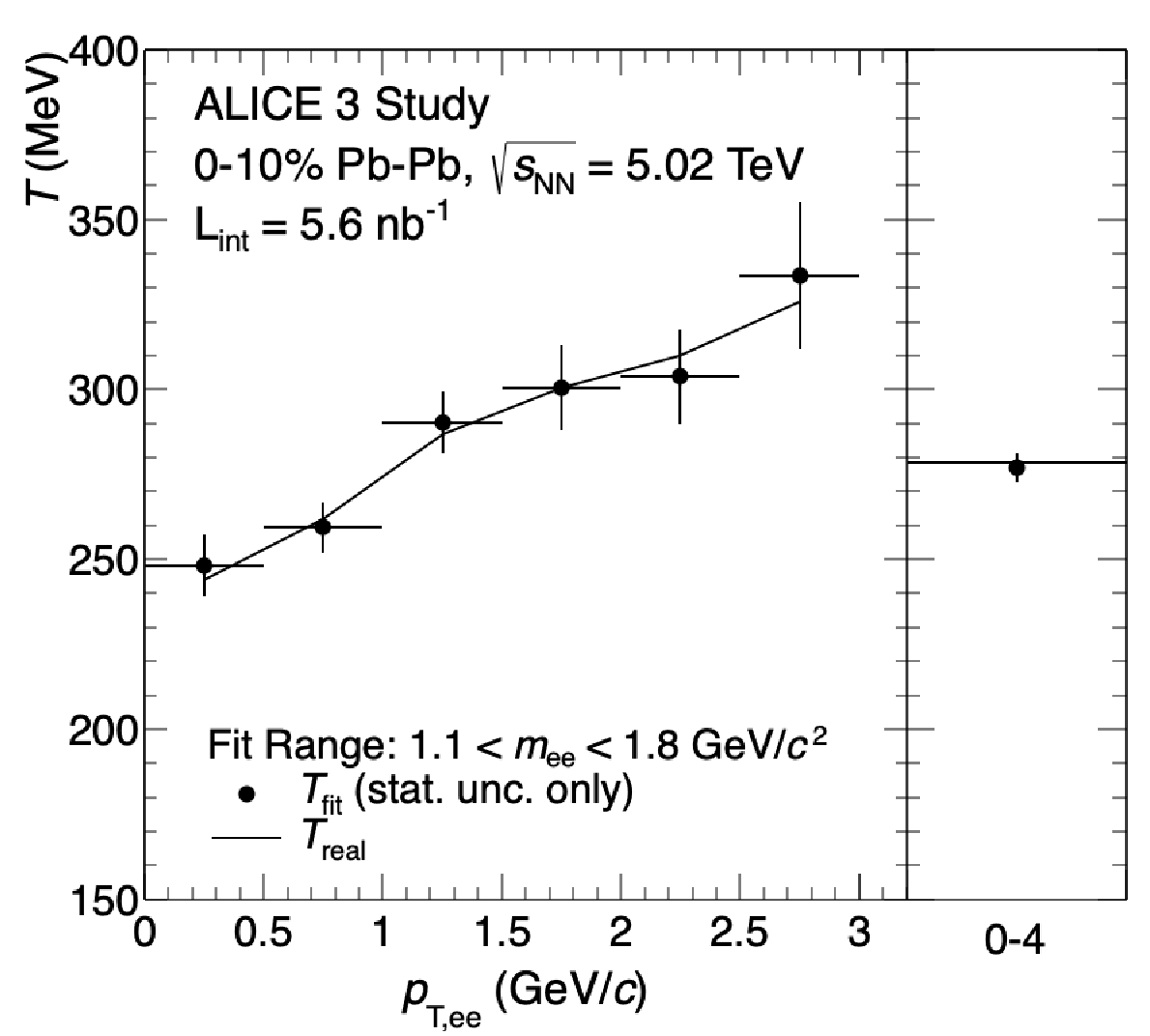}
		\caption{Estimate for the extraction of temperature values ($T$) for different selections on dielectron pair \pT, based on fits over the range $1.1 < m_{\rm ee} < 1.8$ GeV/$c^2$ on simulations.}
		\label{fig:ee_p_T}
	\end{minipage}
\end{figure}
Electromagnetic particles penetrate the strongly interacting medium. Electron pairs provide unique probes to study QCD in the hadronic phase via their dielectron decays as well as to understand the thermodynamical properties of QGP via thermal photons. Its excellent electron identification capabilities down to low \pT and large rapidity coverage makes ALICE 3 an excellent tool to carry out dielectron measurements. The chiral and $U_A(1)$ symmetries may be restored in a hot and dense matter, which manifests in the modification of spectral functions corresponding to vector and pseudoscalar mesons, and influences their spectra~\cite{Rapp:2013nxa,Vargyas:2012ci}. Figure~\ref{fig:ee_lowmass} shows the sensitivity of ALICE 3 dielectron measurements 
to the spectral function modification of the $\rho$ meson in the low-invariant-mass ($m_{\rm ee}\lessapprox 1$ GeV/$c^2$) region. The temperature of the QGP can, on the other hand, be inferred using the precise measurement of the thermal radiation via internal photon conversion ($\gamma^\ast\rightarrow{\rm e}^{+}{\rm e}^{-}$) in the $1\lessapprox m_{\rm ee}\lessapprox 3$ GeV/$c^2$ range~\cite{PHENIX:2008uif}. ALICE 3 will have the unique capability of completing a double-differential measurement in $m_{ee}$ and dielectron pair \pT, and obtain information on the time evolution of thermal properties. Figure~\ref{fig:ee_p_T} quantifies the expected accuracy of the early-time temperature measurement in different pair-\pT ranges, using an estimate of the reconstructed temperature from simulations.

\begin{figure}[h]
	\centering
	\begin{minipage}{.48\textwidth}
		\centering
		\includegraphics[width=\linewidth]{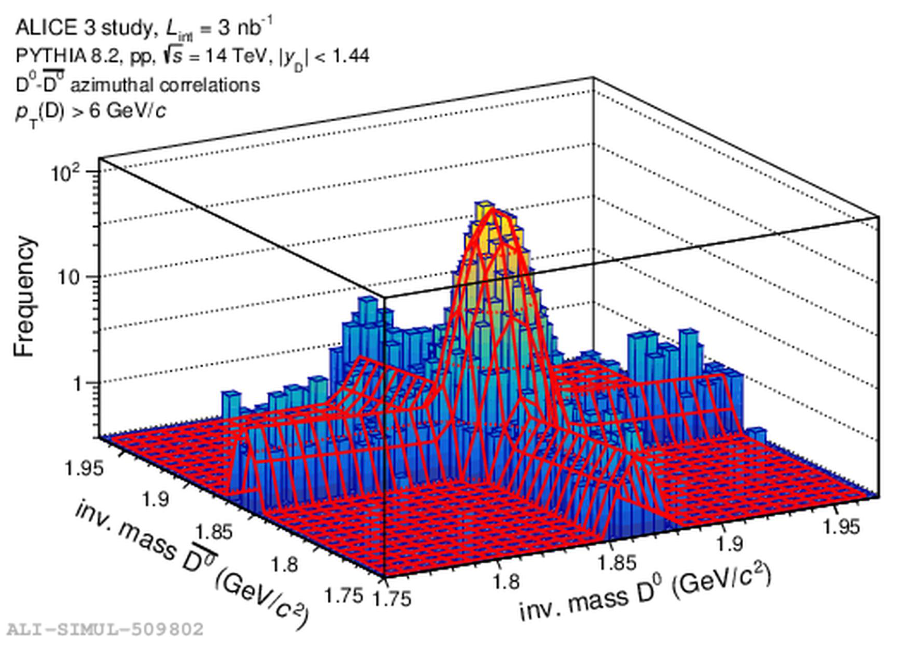}
		\caption{Two-dimensional ${\rm D}^0$--$\overline{{\rm D}}^0$ invariant-mass distributions with $\pT>6$~GeV/$c$, from simulated pp collisions at \sqs=14~TeV, fitted with a Gaussian and a linear background corresponding to both variables.}
		\label{fig:DDinvmass}
	\end{minipage}
	\hfill
	\begin{minipage}{.48\textwidth}
		\centering
		\includegraphics[width=\linewidth]{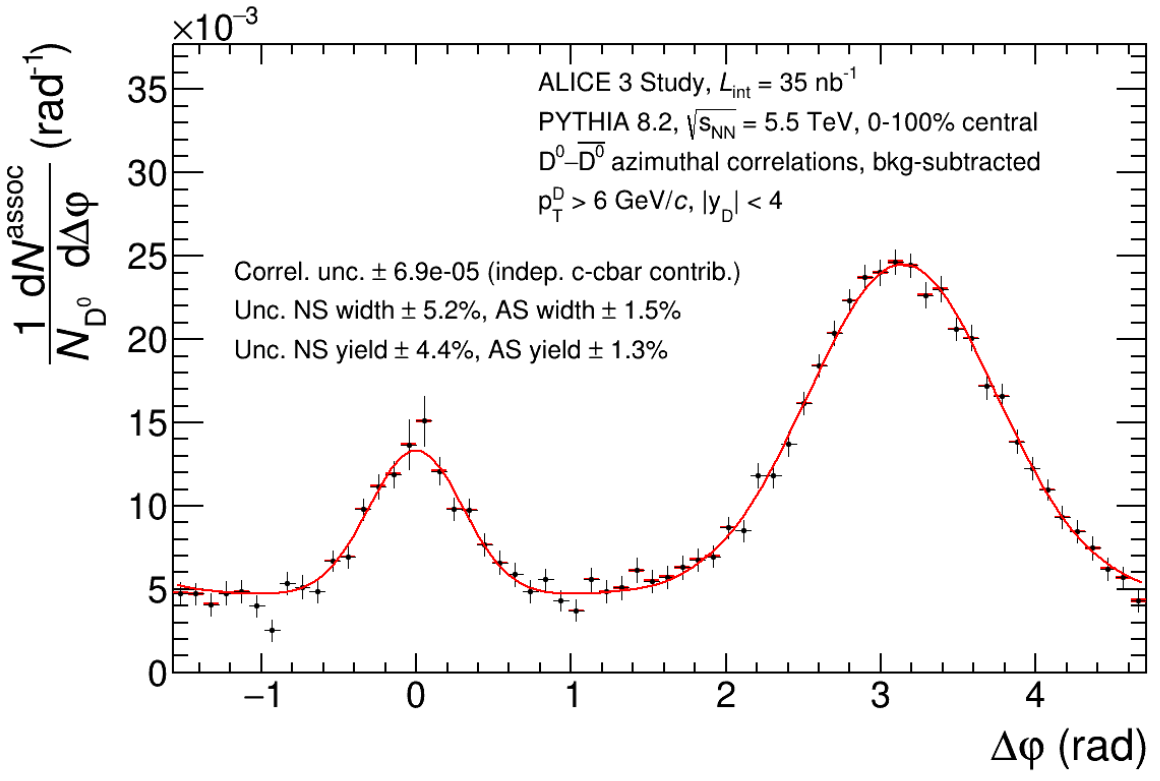}
		\caption{Background-subtracted azimuthal distribution of ${\rm D}^0$--$\overline{{\rm D}}^0$ pairs in Pb--Pb collisions at \sqsn=5.5 TeV with $\pT>6$~GeV/$c$.}
		\label{fig:DDcorr}
	\end{minipage}
\end{figure}
Azimuthal correlations of ${\rm D}^0$--$\overline{{\rm D}}^0$ pairs in pp collisions provide an opportunity to understand the relevance of charm pair-creation, gluon-splitting and flavor-excitation processes that are masked in D--h correlations by fragmentation effects~\cite{Horvath:2023lho}. In Pb--Pb collisions, they provide a direct measure of momentum broadening by the QGP, which is sensitive to the nature of the energy loss mechanisms and to the degree of charm thermalization in the medium~\cite{Nahrgang:2013saa}. Figure~\ref{fig:DDcorr} depicts the yield extraction method using a two-dimensional invariant-mass-distribution fit, while Fig.~\ref{fig:DDcorr} shows the expected precision of the ALICE 3 azimuthal-correlation measurements in heavy-ion collisions corresponding to an integrated luminosity of $\mathcal{L}_{\rm int} = 35$ nb$^{-1}$. 

ALICE 3 will be capable of measuring charm and beauty-meson and baryon production and azimuthal anisotropy (flow) with high precision, shedding light on the degree of flavor-dependent termalization as well as on possible coalescence mechanisms~\cite{Das:2013kea}. Precise measurements of quarkonium excited states will provide an unprecedentedly detailed picture on QGP temperature evolution~\cite{Mocsy:2013syh}. Multi-charm and strange-charm measurements will provide sensitive probes that contribute to the understanding of fragmentation in the yet-underexplored heavy-flavor baryon sector~\cite{Andronic:2021erx}.
\begin{figure}[h]
	\centering
	\begin{minipage}{.48\textwidth}
		\centering
		\includegraphics[width=\linewidth]{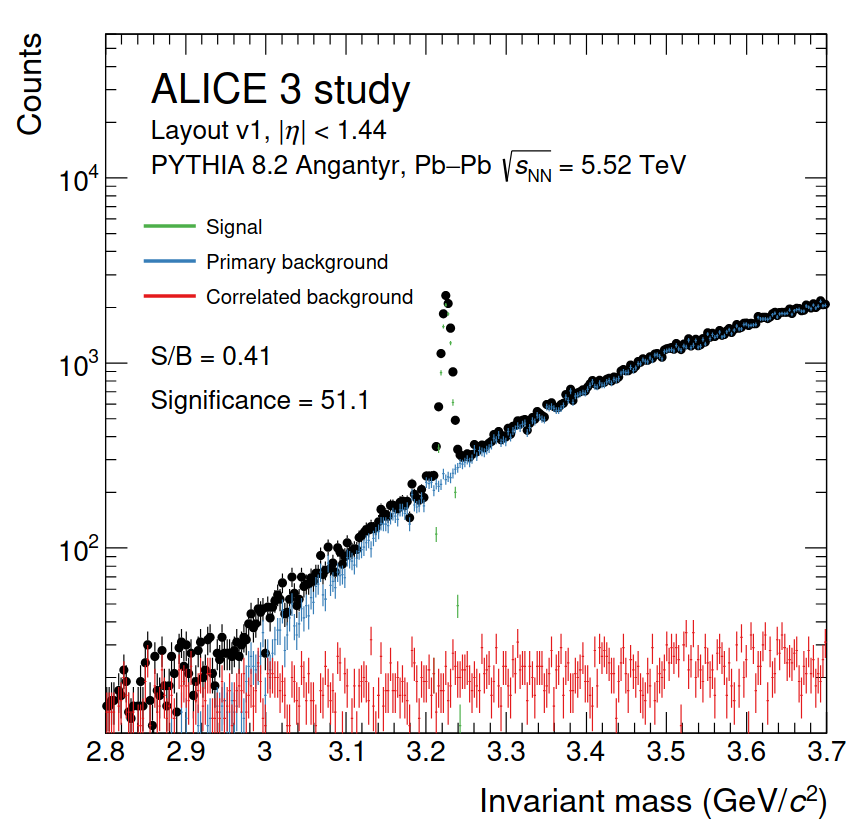}
		\caption{Invariant-mass distribution for the ${\rm c_d} \rightarrow {\rm d}+{\rm K}+\pi$ signal together with correlated background and background from primary deuterons.}
		\label{fig:cd_invmass}
	\end{minipage}
	\hfill
	\begin{minipage}{.48\textwidth}
		\centering
		\includegraphics[width=\linewidth]{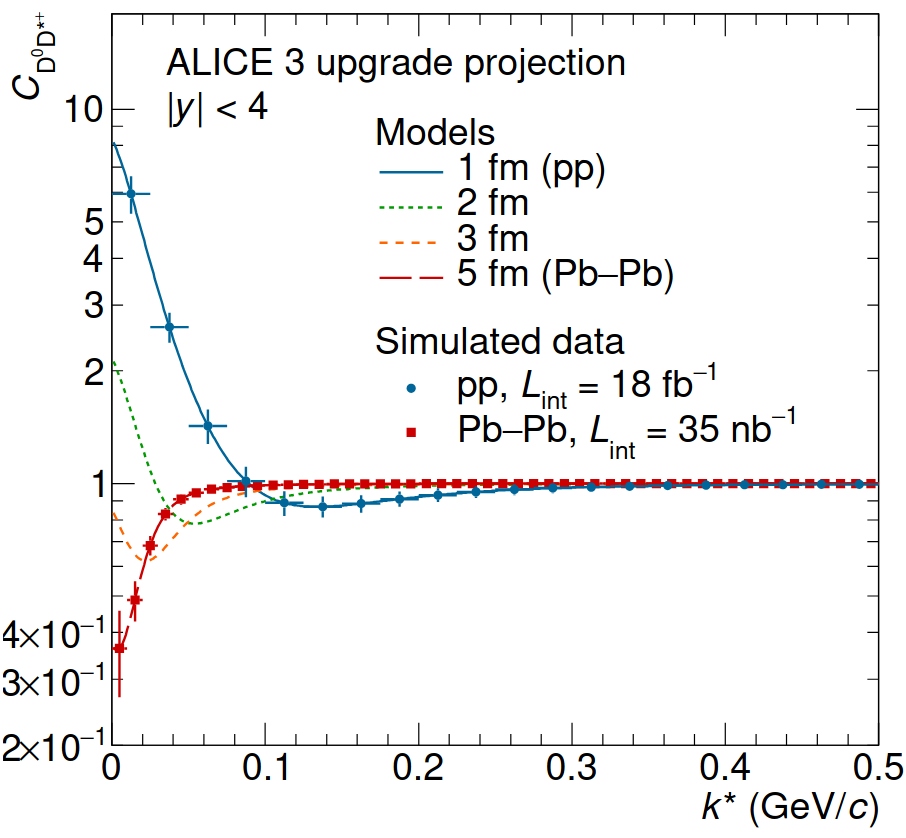}
		\caption{\Dz--\Dsp correlation as a function of the relative momentum $k^\ast$ evaluated for four different source sizes, corresponding to estimates for pp and Pb--Pb systems.}
		\label{fig:DDfemto}
	\end{minipage}
\end{figure}
The extreme precision and high acceptance of ALICE 3 will also make it possible to measure exotic hadronic states incliuding (anti-)hypernuclei, and the first-time observation of supernuclei is also feasible. The most accessible of these are the c-deuteron c$_{\rm d}$ and c-triton c$_{\rm t}$, bound states of a charmed baryon and one or two nucleons ($\Lcp$n and $\Lcp$nn), respectively. Figure~\ref{fig:cd_invmass} shows c$_{\rm d}$ reconstructed via the $\Lcp \rightarrow \pi+{\rm K}+{\rm p}$ channel. The estimated signal corresponds to an integrated luminosity of $\mathcal{L}_{\rm int} = 5.6$ nb$^{-1}$. 
 
Femtoscopic measurements aim to investigate interactions via quantum momentum-correlations. Two-particle femtoscopy of D mesons provide a unique opportunity to test the long range strong interaction. The T$^+_{\rm cc}$ narrow $\Dz \Dz \pi^+$ state has recently been observed by LHCb~\cite{LHCb:2021auc}. However, it is yet unclear whether this state has a tetraquark or a \Dz--\Dsp molecular structure. This question can be resolved by femtoscopic methods. Assuming a molecular structure, the correlation function $C_{\Dz\Dsp}(k^\ast)$ has been calculated, where $k^\ast$ is the relative momentum in the pair rest frame (Fig.~\ref{fig:DDfemto}). 

\section{Summary and outlook}

In conclusion, ALICE 3 is an innovative detector concept that addresses fundamental open questions of quantum-chromodynamics and physics beyond the Standard Model. Its unique capabilities are needed to explore the microscopic dynamics of the quark-gluon plasma, while fully exploiting the capabilities of the high-luminosity LHC. 
To achieve these goals, the physics program planned for ALICE 3 includes the precise measurement of dilepton pairs, heavy-flavor production and correlations, and the reconstruction of quarkonium states as well as exotic hadrons and nuclei, among several other studies.

The currently ongoing steps in detector scoping and preparations consist of wide research efforts such as beam tests and detector simulations. Selection of technologies will be completed and small-scale prototypes will be built until 2025. Large-scale prototypes will be constructed and Technical Design Reports will be completed until 2027. Construction and testing is planned for 2028--2032. Finally, preparations of the cavern and installation will happen until 2034.

\section*{Acknowledgements}

R.V. is grateful for the ALICE colleagues for their help in preparation of this material. This work has been supported by the Hungarian NKFIH OTKA FK131979, OTKA K135515, TKP2021-NKTA-10 and 2021-4.1.2-NEMZ\_KI-2024-00034 grants, and used the infrastructure of the Vesztergombi Laboratory for High Energy Physics (VLAB).



\begin{thebibliography}{12}
	\ifx \bisbn   \undefined \def \bisbn  #1{ISBN #1}\fi
	\ifx \binits  \undefined \def \binits#1{#1}\fi
	\ifx \bauthor  \undefined \def \bauthor#1{#1}\fi
	\ifx \batitle  \undefined \def \batitle#1{#1}\fi
	\ifx \bjtitle  \undefined \def \bjtitle#1{#1}\fi
	\ifx \bvolume  \undefined \def \bvolume#1{\textbf{#1}}\fi
	\ifx \byear  \undefined \def \byear#1{#1}\fi
	\ifx \bissue  \undefined \def \bissue#1{#1}\fi
	\ifx \bfpage  \undefined \def \bfpage#1{#1}\fi
	\ifx \blpage  \undefined \def \blpage #1{#1}\fi
	\ifx \burl  \undefined \def \burl#1{\textsf{#1}}\fi
	\ifx \doiurl  \undefined \def \doiurl#1{\url{https://doi.org/#1}}\fi
	\ifx \betal  \undefined \def \betal{\textit{et al.}}\fi
	\ifx \binstitute  \undefined \def \binstitute#1{#1}\fi
	\ifx \binstitutionaled  \undefined \def \binstitutionaled#1{#1}\fi
	\ifx \bctitle  \undefined \def \bctitle#1{#1}\fi
	\ifx \beditor  \undefined \def \beditor#1{#1}\fi
	\ifx \bpublisher  \undefined \def \bpublisher#1{#1}\fi
	\ifx \bbtitle  \undefined \def \bbtitle#1{#1}\fi
	\ifx \bedition  \undefined \def \bedition#1{#1}\fi
	\ifx \bseriesno  \undefined \def \bseriesno#1{#1}\fi
	\ifx \blocation  \undefined \def \blocation#1{#1}\fi
	\ifx \bsertitle  \undefined \def \bsertitle#1{#1}\fi
	\ifx \bsnm \undefined \def \bsnm#1{#1}\fi
	\ifx \bsuffix \undefined \def \bsuffix#1{#1}\fi
	\ifx \bparticle \undefined \def \bparticle#1{#1}\fi
	\ifx \barticle \undefined \def \barticle#1{#1}\fi
	\bibcommenthead
	\ifx \bconfdate \undefined \def \bconfdate #1{#1}\fi
	\ifx \botherref \undefined \def \botherref #1{#1}\fi
	\ifx \url \undefined \def \url#1{\textsf{#1}}\fi
	\ifx \bchapter \undefined \def \bchapter#1{#1}\fi
	\ifx \bbook \undefined \def \bbook#1{#1}\fi
	\ifx \bcomment \undefined \def \bcomment#1{#1}\fi
	\ifx \oauthor \undefined \def \oauthor#1{#1}\fi
	\ifx \citeauthoryear \undefined \def \citeauthoryear#1{#1}\fi
	\ifx \endbibitem  \undefined \def \endbibitem {}\fi
	\ifx \bconflocation  \undefined \def \bconflocation#1{#1}\fi
	\ifx \arxivurl  \undefined \def \arxivurl#1{\textsf{#1}}\fi
	\csname PreBibitemsHook\endcsname
	
	\bibitem[\protect\citeauthoryear{Acharya et~al.}{2024}]{ALICE:2022wpn}
	\begin{barticle}
		\bauthor{\bsnm{Acharya}, \binits{S.}}, \betal:
		\batitle{{The ALICE experiment: a journey through QCD}}.
		\bjtitle{Eur. Phys. J. C}
		\bvolume{84}(\bissue{8}),
		\bfpage{813}
		(\byear{2024})
		\doiurl{10.1140/epjc/s10052-024-12935-y}
		{\href{https://arxiv.org/abs/2211.04384}{{arXiv:2211.04384}}}
		{[nucl-ex]}
	\end{barticle}
	\endbibitem
	
	\bibitem[\protect\citeauthoryear{}{2022}]{ALICE:2022wwr}
	\begin{botherref}
		{Letter of intent for ALICE 3: A next-generation heavy-ion experiment at the
			LHC}
		(2022)
		{\href{https://arxiv.org/abs/2211.02491}{{arXiv:2211.02491}}}
		{[physics.ins-det]}
	\end{botherref}
	\endbibitem
	
	\bibitem[\protect\citeauthoryear{Alfaro et~al.}{2024}]{Alfaro:2024sxc}
	\begin{barticle}
		\bauthor{\bsnm{Alfaro}, \binits{R.}}, \betal:
		\batitle{{Characterisation of plastic scintillator paddles and lightweight
				MWPCs for the MID subsystem of ALICE 3}}.
		\bjtitle{JINST}
		\bvolume{19}(\bissue{04}),
		\bfpage{04006}
		(\byear{2024})
		\doiurl{10.1088/1748-0221/19/04/T04006}
		{\href{https://arxiv.org/abs/2401.04630}{{arXiv:2401.04630}}}
		{[physics.ins-det]}
	\end{barticle}
	\endbibitem
	
	\bibitem[\protect\citeauthoryear{Rapp}{2013}]{Rapp:2013nxa}
	\begin{barticle}
		\bauthor{\bsnm{Rapp}, \binits{R.}}:
		\batitle{{Dilepton Spectroscopy of QCD Matter at Collider Energies}}.
		\bjtitle{Adv. High Energy Phys.}
		\bvolume{2013},
		\bfpage{148253}
		(\byear{2013})
		\doiurl{10.1155/2013/148253}
		{\href{https://arxiv.org/abs/1304.2309}{{arXiv:1304.2309}}}
		{[hep-ph]}
	\end{barticle}
	\endbibitem
	
	\bibitem[\protect\citeauthoryear{Vargyas et~al.}{2013}]{Vargyas:2012ci}
	\begin{barticle}
		\bauthor{\bsnm{Vargyas}, \binits{M.}},
		\bauthor{\bsnm{Cs\"org\H{o}}, \binits{T.}},
		\bauthor{\bsnm{V\'ertesi}, \binits{R.}}:
		\batitle{{Effects of chain decays, radial flow and $U_{A}(1)$ restoration on
				the low-mass dilepton enhancement in $\sqrt{s_{NN}}=200$ GeV Au+Au
				reactions}}.
		\bjtitle{Central Eur. J. Phys.}
		\bvolume{11},
		\bfpage{553}--\blpage{559}
		(\byear{2013})
		\doiurl{10.2478/s11534-013-0249-6}
		{\href{https://arxiv.org/abs/1211.1166}{{arXiv:1211.1166}}}
		{[nucl-th]}
	\end{barticle}
	\endbibitem
	
	\bibitem[\protect\citeauthoryear{Adare et~al.}{2010}]{PHENIX:2008uif}
	\begin{barticle}
		\bauthor{\bsnm{Adare}, \binits{A.}}, \betal:
		\batitle{{Enhanced production of direct photons in Au+Au collisions at
				$\sqrt{s_{NN}}=200$ GeV and implications for the initial temperature}}.
		\bjtitle{Phys. Rev. Lett.}
		\bvolume{104},
		\bfpage{132301}
		(\byear{2010})
		\doiurl{10.1103/PhysRevLett.104.132301}
		{\href{https://arxiv.org/abs/0804.4168}{{arXiv:0804.4168}}}
		{[nucl-ex]}
	\end{barticle}
	\endbibitem
	
	\bibitem[\protect\citeauthoryear{Horv\'ath et~al.}{2023}]{Horvath:2023lho}
	\begin{barticle}
		\bauthor{\bsnm{Horv\'ath}, \binits{A.}},
		\bauthor{\bsnm{Frajna}, \binits{E.}},
		\bauthor{\bsnm{V\'ertesi}, \binits{R.}}:
		\batitle{{Event-Shape-Dependent Analysis of Charm\textendash{}Anticharm
				Azimuthal Correlations in Simulations}}.
		\bjtitle{Universe}
		\bvolume{9}(\bissue{7}),
		\bfpage{308}
		(\byear{2023})
		\doiurl{10.3390/universe9070308}
		{\href{https://arxiv.org/abs/2306.05910}{{arXiv:2306.05910}}}
		{[hep-ph]}
	\end{barticle}
	\endbibitem
	
	\bibitem[\protect\citeauthoryear{Nahrgang et~al.}{2014}]{Nahrgang:2013saa}
	\begin{barticle}
		\bauthor{\bsnm{Nahrgang}, \binits{M.}},
		\bauthor{\bsnm{Aichelin}, \binits{J.}},
		\bauthor{\bsnm{Gossiaux}, \binits{P.B.}},
		\bauthor{\bsnm{Werner}, \binits{K.}}:
		\batitle{{Azimuthal correlations of heavy quarks in Pb + Pb collisions at
				$\sqrt{s}=2.76$ TeV at the CERN Large Hadron Collider}}.
		\bjtitle{Phys. Rev. C}
		\bvolume{90}(\bissue{2}),
		\bfpage{024907}
		(\byear{2014})
		\doiurl{10.1103/PhysRevC.90.024907}
		{\href{https://arxiv.org/abs/1305.3823}{{arXiv:1305.3823}}}
		{[hep-ph]}
	\end{barticle}
	\endbibitem
	
	\bibitem[\protect\citeauthoryear{Das et~al.}{2014}]{Das:2013kea}
	\begin{barticle}
		\bauthor{\bsnm{Das}, \binits{S.K.}},
		\bauthor{\bsnm{Scardina}, \binits{F.}},
		\bauthor{\bsnm{Plumari}, \binits{S.}},
		\bauthor{\bsnm{Greco}, \binits{V.}}:
		\batitle{{Heavy-flavor in-medium momentum evolution: Langevin versus Boltzmann
				approach}}.
		\bjtitle{Phys. Rev. C}
		\bvolume{90},
		\bfpage{044901}
		(\byear{2014})
		\doiurl{10.1103/PhysRevC.90.044901}
		{\href{https://arxiv.org/abs/1312.6857}{{arXiv:1312.6857}}}
		{[nucl-th]}
	\end{barticle}
	\endbibitem
	
	\bibitem[\protect\citeauthoryear{M\'ocsy et~al.}{2013}]{Mocsy:2013syh}
	\begin{barticle}
		\bauthor{\bsnm{M\'ocsy}, \binits{A.}},
		\bauthor{\bsnm{Petreczky}, \binits{P.}},
		\bauthor{\bsnm{Strickland}, \binits{M.}}:
		\batitle{{Quarkonia in the Quark Gluon Plasma}}.
		\bjtitle{Int. J. Mod. Phys. A}
		\bvolume{28},
		\bfpage{1340012}
		(\byear{2013})
		\doiurl{10.1142/S0217751X13400125}
		{\href{https://arxiv.org/abs/1302.2180}{{arXiv:1302.2180}}}
		{[hep-ph]}
	\end{barticle}
	\endbibitem
	
	\bibitem[\protect\citeauthoryear{Andronic et~al.}{2021}]{Andronic:2021erx}
	\begin{barticle}
		\bauthor{\bsnm{Andronic}, \binits{A.}},
		\bauthor{\bsnm{Braun-Munzinger}, \binits{P.}},
		\bauthor{\bsnm{K\"ohler}, \binits{M.K.}},
		\bauthor{\bsnm{Mazeliauskas}, \binits{A.}},
		\bauthor{\bsnm{Redlich}, \binits{K.}},
		\bauthor{\bsnm{Stachel}, \binits{J.}},
		\bauthor{\bsnm{Vislavicius}, \binits{V.}}:
		\batitle{{The multiple-charm hierarchy in the statistical hadronization
				model}}.
		\bjtitle{JHEP}
		\bvolume{07},
		\bfpage{035}
		(\byear{2021})
		\doiurl{10.1007/JHEP07(2021)035}
		{\href{https://arxiv.org/abs/2104.12754}{{arXiv:2104.12754}}}
		{[hep-ph]}
	\end{barticle}
	\endbibitem
	
	\bibitem[\protect\citeauthoryear{Aaij et~al.}{2022}]{LHCb:2021auc}
	\begin{barticle}
		\bauthor{\bsnm{Aaij}, \binits{R.}}, \betal:
		\batitle{{Study of the doubly charmed tetraquark $T_{cc}^{+}$}}.
		\bjtitle{Nature Commun.}
		\bvolume{13}(\bissue{1}),
		\bfpage{3351}
		(\byear{2022})
		\doiurl{10.1038/s41467-022-30206-w}
		{\href{https://arxiv.org/abs/2109.01056}{{arXiv:2109.01056}}}
		{[hep-ex]}
	\end{barticle}
	\endbibitem
	
\end{thebibliography}

\end{document}